\documentclass[aps,prstab,twocolumn,superscriptaddress]{revtex4-1}

\usepackage{amsmath}
\usepackage{amssymb}
\usepackage{mathtools}

\usepackage{booktabs}
\usepackage{graphicx}
\usepackage{color}
\usepackage[caption=false]{subfig}

\usepackage[english]{babel}
\usepackage{bm}
\usepackage{braket}

\begin{document}

\title{A muon source based on planar channelling radiation}


\author{X. Buffat}
\email{xavier.buffat@cern.ch}
\affiliation{European Organization for Nuclear Research (CERN), CH-1211 Geneva, Switzerland}



\date{\today}

\begin{abstract}
The usage of channelling radiation arising from a high energy electron beam traversing a crystal for the production of an intense low emittance muon beam is investigated. The optimal energy and divergence of the electron beam are computed for few crystal types based on analytical expressions of the radiation spectrum and of the muon pair production cross section. It is shown that the required beam properties seem within reach and that the source performance may be comparable to other muon sources currently considered.
\end{abstract}


\maketitle

\section{Introduction}
The radiation emitted by high energy particles trapped transversely in the potential generated by a crystalline lattice aligned with their direction of travel, so-called channelling radiation, have been extensively studied mostly aiming at the generation of high energy and high brightness photon beams for various applications, e.g. in~\cite{channellingBook} and references therein. In the following, we investigate the potential of such a source of high energy photons for the generation of intense low emittance muon beams, via pair production on a target. The model for the channelling radiation is described based on existing literature in the first section, along with the interaction of the photon beam with the target generating the positive and negative muon beams. This allows to determine optimal parameters for the beam traversing the crystal. The zero order design of the electron accelerator meeting the requirement of this scheme is discussed in the second section, allowing for an estimation of the potential performance and its comparison to other types of sources.
\section{Channelling radiation} \label{ref-channel}
\begin{figure}
\begin{center}
 \subfloat[Oscillation amplitude (dashed lines) and undulator parameters (solid lines).]{\includegraphics[width=0.9\linewidth]{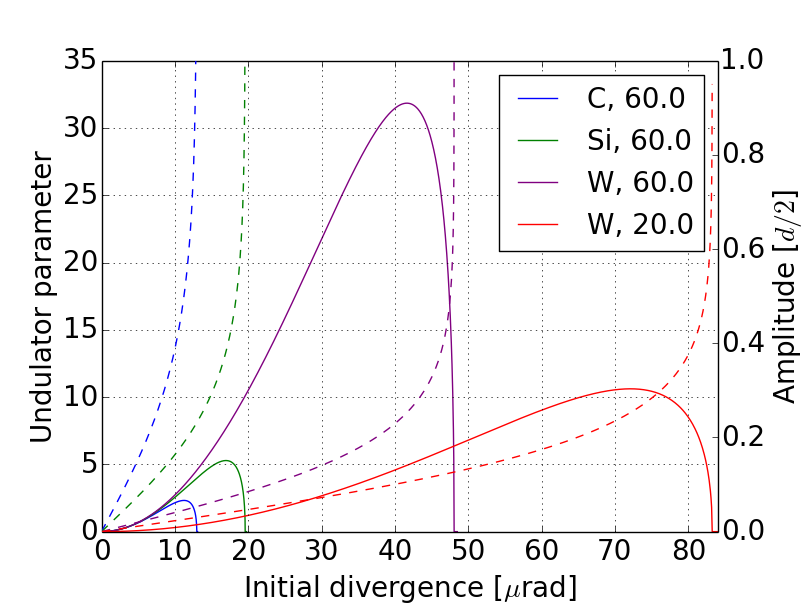}\label{fig-undulator}}
 \qquad 
 \subfloat[Photon energy (dashed lines) and rate (solid lines).]{\includegraphics[width=0.9\linewidth]{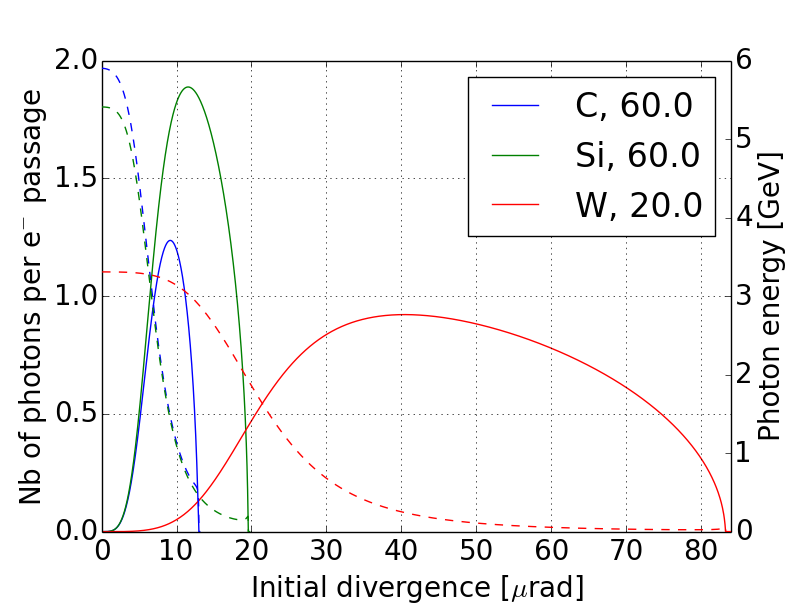}\label{fig-undulator-EnergyAndRate}}
 \end{center}
\caption{Properties of the electron trajectory in the crystal (upper plot) and of the emitted photons (lower plot) as a function of the divergence of an electron entering a crystal at the centre of a channel. The crystal types and electron energies are listed in the legend, the corresponding parameter of the PT potential for the various crystals can be found in Tab.~\ref{tab-crystal}.}
\end{figure}
\begin{figure}
 \begin{center}
  \includegraphics[width=0.9\linewidth]{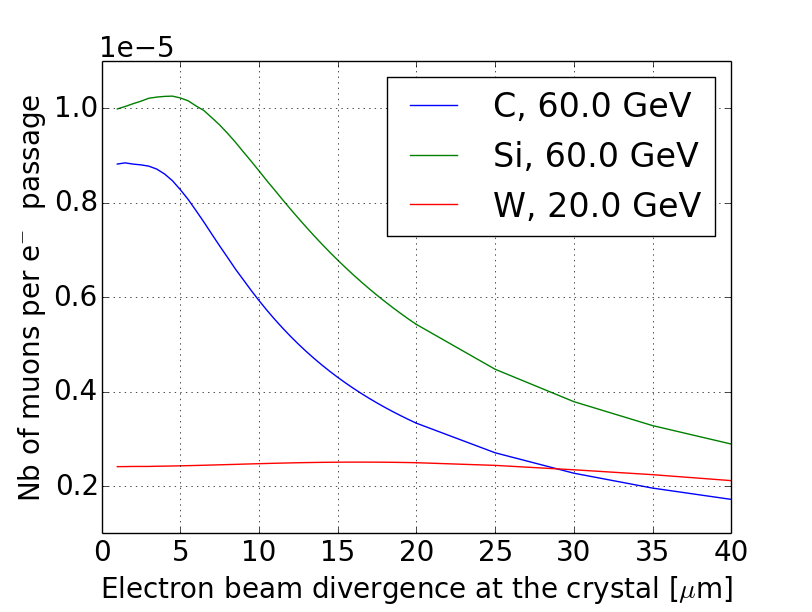}
 \end{center}
\caption{Number of muons produced by the passage of an electron through the crystal within an energy acceptance of $\pm10\%$ as a function of the electron beam divergence at the crystal. The crystal types and electron energies in GeV are listed in the legend, the corresponding parameter of the PT potential for the various crystals can be found in Tab.~\ref{tab-crystal}. The crystal length is set to the dechannelling length (Tab.~\ref{tab-source}).}
\label{fig-optimalDivergence}
\end{figure}
\begin{figure}
 \begin{center}
   \subfloat{\includegraphics[width=0.9\linewidth]{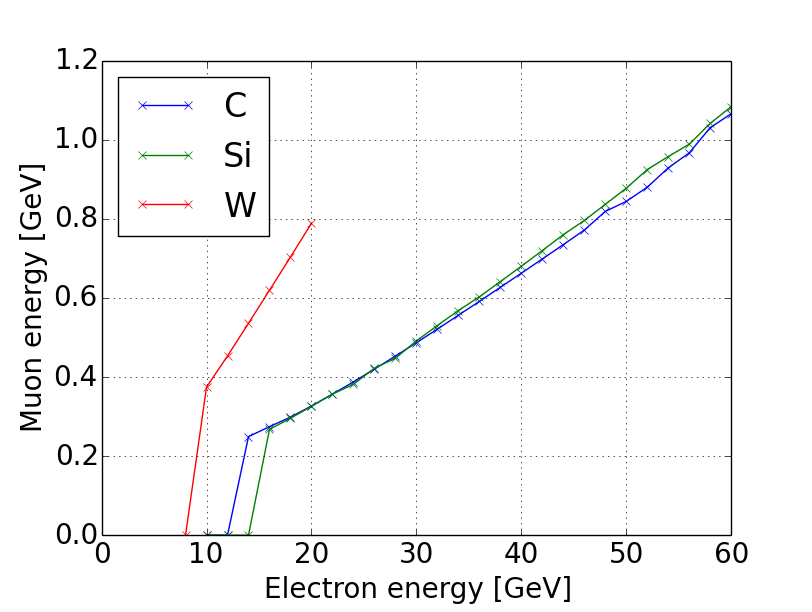}}
  \qquad
  \subfloat{\includegraphics[width=0.9\linewidth]{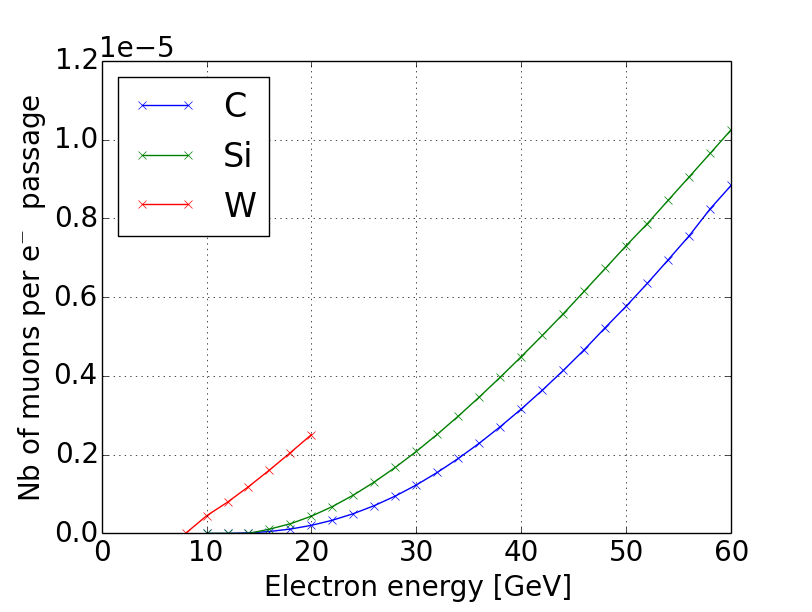}}
 \end{center}
\caption{Muon output energy and rate per electron passage through the crystal as a function of the electron beam energy for carbon, silicon and tungsten crystals.}
\label{fig-optimalEnergy}
\end{figure}
\begin{figure}
\begin{center}
\includegraphics[width=0.9\linewidth]{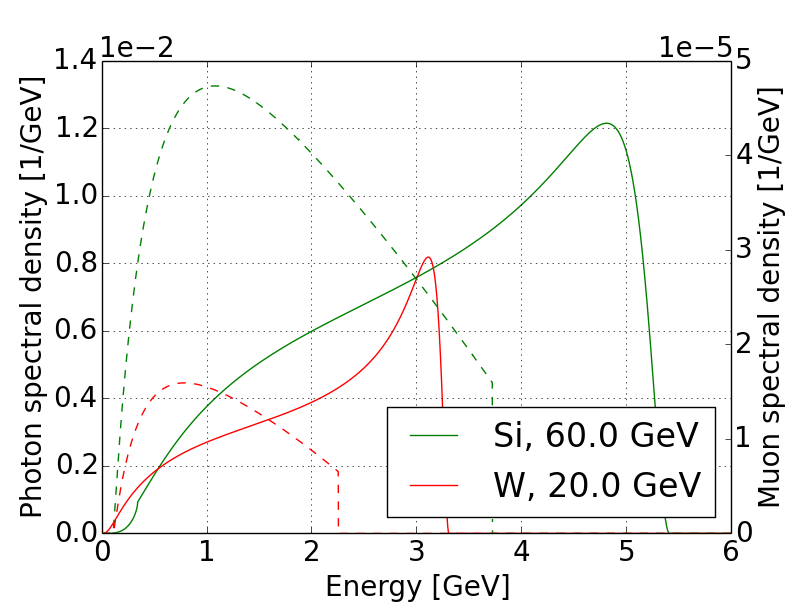}
\end{center}
 \caption{Spectral density of the photons (solid line) and muons (dashed line) generated by the passage of an electron through the crystal for the two designs described in Tab.~\ref{tab-source}.}
 \label{fig-photonMuonSpectrum}
\end{figure}

Following ~\cite{channellingBook}, we derive the radiation spectrum of channelled high energy electrons in a crystal potential by considering stable trajectories in the P\"oschl-Teller (PT) potential and integrating the emitted spectrum of individual electrons over the incoming beam phase space. Typical PT potential parameters $a_{PT}$ and $b_{PT}$ as well as the inter-planar distance $d$ and the height of the channelling potential $U_c$ are listed in Tab.~\ref{tab-crystal}. Since we are targeting high energy electrons, the classical treatment is appropriate. The PT potential at a given distance from the mid-plane $x$ reads:
\begin{equation}
 U_{PT}(x) = a_{PT}\tanh^2\left(\frac{x}{b_{PT}}\right).
\end{equation}
An incoming electron at $x$ with an energy $E_e$ and an angle $x'$ with respect to the channelling plane has an oscillatory trajectory in the potential characterised by an amplitude
\begin{equation}
A_e = b_{PT}\tanh^{-1}\sqrt{\frac{E_ex'^2+U_{PT}(x)}{a_{PT}}},
\end{equation}
and a wave number
\begin{equation}
 k_u = \sqrt{\frac{2a_{PT}}{E_e}}\frac{1}{b_{PT}\cosh\left(\frac{A_e}{b_{PT}}\right)}.
\end{equation}
The corresponding undulator parameter is given by:
\begin{equation}
 K^2 = 4\gamma_e\frac{a_{PT}}{m_ec^2}\frac{\cosh\left(\frac{A_e}{b_{PT}}\right)-1}{\cosh^2\left(\frac{A_e}{b_{PT}}\right)},
\end{equation}
with $m_e$ the mass of the electron and $c$ the speed of light. The amplitude and the corresponding undulator parameter for electrons with different coordinates at the entrance of the crystal are shown in Fig.~\ref{fig-undulator}. The tungsten crystal, featuring a significantly higher potential, can channel particles with a higher amplitude, therefore larger undulator parameter. The channelled particles remain mostly in the undulator regime, i.e. $K\approx1$, only for low electron energies. In this condition we expect a narrow photon energy spectrum in the first harmonic of the undulator. In the following we will discard the energy lost in higher harmonics. Since this approximation is not valid for large undulator parameters, we will not consider electron beams higher than 20 GeV with the tungsten crystal. Also, since we are aiming at the energy spectrum of the muons, for which the pair production spectrum is wide, we shall approximate the undulator spectrum of each electron by a delta function peaked at~\cite{undulatorTheory}:
\begin{equation}
E_{\gamma}=\frac{2\hbar c\gamma_e^2k_u}{1+\frac{K^2}{2}} \label{eq-photonEnergy}
\end{equation}
with $\hbar$ the reduced Plank constant. The number of photons emitted per electron and per passage through a crystal of length $L_c$ is given by:
\begin{equation}
 n_{\gamma} = \frac{e^2k_u}{24\pi\epsilon_0\hbar}\frac{K^2}{(1+K^2/2)^2}\frac{L_c}{c}.\label{eq-photonNumber}
\end{equation}
The crystal length is set equal to the dechannelling length, given by~\cite{channellingBook}:
\begin{equation}
 L_{d-c} = 8.9\cdot10^{-6}U_cE_eL_r,
\end{equation}
with the radiation length $L_r$. The corresponding values for different designs are shown in Tab.~\ref{tab-source}. The energy lost by an electron when traversing the crystal can be obtained using Eqs.~\ref{eq-photonEnergy} and~\ref{eq-photonNumber}. Its maximum does not exceed 10\% of the initial electron energy in the designs considered here, such that the approximation of constant energy is valid. \\
The photon energy and rate expected for particles with different coordinates at the entrance of the crystal are shown in Fig.~\ref{fig-undulator-EnergyAndRate}. The photons with highest energy are produced by low amplitude particles, however due to the low undulator parameter the corresponding rates are low. On the other hand, large amplitude particles produce high rates of low energy photons. An optimisation of the electron beam divergence is therefore required to obtain the highest rates in a given range of interest. \\
Considering a Gaussian beam of r.m.s. divergence $\sigma'_e$ and r.m.s. beam size $\sigma_e$ much larger than the distance between crystalline planes $d$, we can write the total spectrum of the radiation by averaging over the beam distribution:
\begin{equation}
 N_\gamma(E') = \frac{1}{d}\int\limits_{-d/2}^{d/2}dx\int\limits_{-\infty}^\infty dx'\frac{n_\gamma}{\sqrt{2\pi\sigma_e'^2}}e^{-\frac{x'^2}{2\sigma_e'^2}}\delta(E'-E_\gamma)
\end{equation}
We shall consider photon energies significantly higher than the muon pair production threshold, thus, given the target atomic number $Z$ and mass number $A$ the differential cross section is given by~\cite{helmut}:
\begin{equation}
 \frac{d\sigma}{dE_{\mu}} = 4\frac{\alpha Z^2r_\mu^2}{E_\gamma}\left(1-\frac{4}{3}\frac{E_\mu}{E_\gamma}\left(1-\frac{E_\mu}{E_\gamma}\right)\right)\log{W}
\end{equation}
with $r_\mu$ the classical radius of the muon, $\alpha$ the fine structure constant and 
\begin{equation}
 W=\frac{BZ^{-1/3}}{D_n}\frac{m_\mu}{m_e}\frac{1+\left(D_n\sqrt{e}-2\right)\frac{\delta}{m_\mu}}{1+BZ^{-1/3}\sqrt{e}\frac{\delta}{m_e}},
\end{equation}
where $e$ is the Euler number, $m_\mu$ the muon mass, $m_e$ the electron mass, $B=183$ and $D_n=1.54A^{0.27}$ and finally:
\begin{equation}
 \delta = \frac{m_\mu^2}{2E_\mu}\left(1-\frac{E_\mu}{E_\gamma}\right).
\end{equation}
We then find the muon energy spectrum by integrating over the photon energies $E'$:
\begin{equation}
 N_\mu(E) = \int_0^\infty N_\gamma(E')\frac{d\sigma_{\mu}}{dE}(E,E')dE'
\end{equation}
This energy spectrum is very wide, we are interested in the fraction of muons that can be injected in a beam line with a finite relative energy acceptance given by $A_E$:
\begin{equation}
 N_\mu(A_E) = \frac{\rho_tL_tN_A}{M_t}\int\limits_{E_{\mu,max}(1-A_E)}^{E_{\mu,max}(1+A_E)}N_\mu(E)dE,
\end{equation}
with $E_{\mu,max}$ the muon energy corresponding to the peak of the spectrum, $L_t$, $\rho_t$ and $M_t$ the length, density and molar mass of the target and $N_A$ the Avogadro number. In the following, we shall consider a tungsten target at room temperature. Its length is chosen below the radiation length, such that the development of the electromagnetic shower is limited. A more accurate estimate of the optimal target length should be obtained with a detailed shower development model. Figure~\ref{fig-optimalDivergence} shows the variation of the muon rate in a given energy acceptance for different electron beam divergence, allowing for an estimation of the optimal divergence in the various configuration. We note that, as the channelling potential of tungsten is higher, the optimal divergence is significantly higher than the other crystals. The maximum rate obtained based on such an optimisation as a function of the electron beam energy is shown in Fig.~\ref{fig-optimalEnergy}. The tungsten crystal outperforms the others in term of number of muons produced per electron passage as well as in terms of output muon energy, but the undulator approximation used in the derivation of the spectrum does not allow for predictions above approximatively 20 GeV. The carbon crystal does not feature any advantages in the energy range considered. The silicon crystal is therefore an interesting option at high energy. In the following, we focus on a low energy option based on a tungsten crystal and a high energy option based on a silicon crystal, the corresponding parameters are listed in Tab.~\ref{tab-source} and the photon and muon spectrum are shown in Fig.~\ref{fig-photonMuonSpectrum}. \\

As the crystal is thin and the photon emission angle narrow, the photon beam properties are mostly identical to the one of the electron beam at the crystal, except for the divergence in the channelling plane which is strongly reduced thanks to the undulation of the electrons in the channel~\cite{undulatorTheory2}. Electron beams with transverse emittances in the order of a few micrometers may be expected using for example Energy Recovery Linear accelerators (ERL) at such energies \cite{eRHIC,LHeC}. Thus the muon beam emittance after the target is mostly dominated by the required length of the target $L_t$ and the emission angle of the muon pair. Indeed, simplifying the dynamics of the pair production by considering that the emission of muons is uniform along the beam trajectory in the target and uniformly distributed within a cone with opening angle $1/\gamma_\mu$, we find that the r.m.s. muon beam emittance is approximatively given by:
\begin{equation}
\epsilon_\mu = \frac{1}{\sqrt{3}}\frac{L_t}{\gamma_\mu^2}.
\end{equation}
The corresponding transverse emittance obtained for the two designs are listed in Tab.~\ref{tab-source}. We note that the electron beam's optical function in the channelling plane $\beta_{\parallel}$ is set according to the optimal divergence at the crystal, whereas in the perpendicular plane $\beta_{\perp}$ is set to minimise the photon beam size at the target based its distance to the crystal.
\begin{table}
 \begin{tabular}{lccc}
 Crystal (plane) & C(100) & Si(110) & W (110) \\
 \hline
 $d$ [\r{A}]  & 0.89 & 1.92 & 2.24 \\
 $a_{PT}$ [eV] & 10.1 & 23.0 & 138.6 \\
 $b_{PT}/d$ [\r{A}] & 0.19 & 0.145 & 0.096 \\
 $U_c$ [eV] & 9.9  & 22.9 & 138.6 \\
 \hline
\end{tabular}
\caption{PT potential parameters and channelling potential height of carbon, silicon and tungsten crystals~\cite{channellingBook}.}
\label{tab-crystal}
\end{table}
\section{Electron beam}\label{ref-beam}
\begin{table}
\begin{tabular}{lcc}
 Electron beam & &  \\
 \hline
 Energy [GeV] & 20 & 60 \\
 Current [mA] & 1.5 & 9.5 \\
 Optimal divergence [$\mu$rad] & 16.5 & 4.5 \\
 Norm. transverse emit. [$\mu$m]& 5 & 50 \\
 $\beta_{\parallel}$ at the crystal [m] & 0.5 & 21.0 \\
 $\beta_{\perp}$ at the crystal [m] & \multicolumn{2}{c}{10.0} \\
 Maximum relative energy loss & \multicolumn{2}{c}{0.1} \\
 \hline
 & \\
 Crystal &\\
 \hline
 Crystal Type  & W(110) & Si (110) \\
 Radiation Length [cm] & 0.35 & 9.37   \\
 Crystal length [mm] & 0.0 & 1.1   \\
 \hline
  & \\
 Target & &  \\
 \hline
 Material & \multicolumn{2}{c}{W} \\
 Length [cm] & \multicolumn{2}{c}{5} \\
 Distance to the crystal [m] & \multicolumn{2}{c}{10} \\
 \hline
 & \\
 Muon beams & &  \\
 \hline
 Energy acceptance & \multicolumn{2}{c}{$\pm10\%$} \\
 Energy [GeV] & 2.0 & 1.1 \\
 Efficiency [$10^{-5}\mu$/electron] & 0.2 & 1.0 \\
 Rate [$10^{11}\mu/s$] & 0.2 & 9.5 \\
 Phys. transverse emit. [mm] &  0.08 & 0.3 \\
 Norm. transverse emit. [mm] &  1.5 & 2.8 \\
 \hline
\end{tabular}
\caption{Muon source parameters. The symbols $\parallel$ and $\perp$ refer to the plane parallel and perpendicular to the channelling plane respectively.}
\label{tab-source}
\end{table}
Let us consider an electron beam circulating in a synchrotron. Due to multiple scattering, the emittance of the circulating beam is significantly affected by the presence of a crystal. We can write the r.m.s. deflection angle~\cite{multipleScatteringMohl}:
\begin{equation}
 \theta_c = \frac{13.6\cdot10^6}{eE_e}\sqrt{\frac{L_c}{L_r}}
\end{equation}
resulting in an emittance growth at each passage through the crystal and, since this contribution is significantly larger than other sources of emittance growth such as quantum excitations, we can write the equilibrium divergence:
\begin{equation}
 \sigma'^2_{e,equ} = \frac{1}{2}\tau_e\theta_c^2
\end{equation}
with $\tau_e$ the damping time of the transverse emittances due to the emission of synchrotron radiation along the ring~\cite{wolski}. Imposing that the equilibrium divergence matches the optimal divergence $\sigma'_{e,opt}$ obtained above, we find the optimal damping time~:
\begin{equation}
 \tau_{e,opt} = 2\frac{\sigma'^2_{e,opt}}{\theta^2_c}.
\end{equation}
For both designs in Tab.~\ref{tab-source}, the required damping time is about half a turn, which is out of reach. As shown in Fig.~\ref{fig-optimalDivergence}, operating with a higher divergence, compatible with longer damping times, would compromise significantly the performance of the source. To overcome the limitation caused by multiple scattering in the crystal over multiple turns, we shall consider single pass technologies, i.e. using a linear accelerator or an ERL. In the following, we will not differentiate the two options since the principle of the source is identical, we note nevertheless that the ERL deceleration requires a sufficiently large energy acceptance, as some electrons might not lose energy while others will lose up to few percent of their energy due to the emission of radiation in the crystal, thus leading to a large spread. \\
Table~\ref{tab-source} shows the estimation of the performance of two designs of muon sources based on tungsten and silicon crystals. The energy and emittance of the electron beam for the tungsten option are comparable to the high power ERL design for the eRHIC project~\cite{eRHIC}, yet featuring a reduced current as discussed below. Similarly, the electron beam parameters for the silicon based option are comparable to those of the  ERL design within the LHeC project~\cite{LHeC}. \\

At high energies the main interaction of the electrons in the beam with the crystal is electron-positron pair production in the Coulombian field of the nuclei. However, the pair-production does not lead to a local deposition of energy, such that this mechanism does not threaten the integrity of the crystal. The electron beam will therefore deposit energy in the crystal mainly via collision with the electrons. This contribution is rather independent of the energy in the range of interest, corresponding to 1.6 and 2.1 MeV / cm$^2$/g for silicon and tungsten~\cite{electronicLossTable}. The currents quoted in Tab.~\ref{tab-source} are such that the power deposited in the crystal is about 50~W, based on the materials density at room temperature. \\
We observe that the silicon crystal is more performing in terms of muon rate not only thanks to the shape of the potential, as discussed in previous section, but also because its low density allows for a high electron beam current for the given energy deposition. A more accurate estimate of the maximum electron beam current should take into account the resistance of the material to thermal load.
\section{Conclusion}
Considering the classical motion of high energy electrons in planar crystals and the muon pair production of the resulting photon beam impacting on a target, it is shown that a realistic electron LINAC or ERL of energy in the tens of GeV could be used to generate muon beams with a rate comparable to pion decay-based sources, yet featuring a smaller transverse emittance~(e.g. \cite{prism,nustorm}). Such sources could be considered for various particle physics applications such as the generation of intense neutrino beams or multi-TeV lepton colliders.
\section{Acknowledgements}
I would like to thank E. M\'etral and G. Arduini for their comments on this work as well as D. Schulte for valuable discussions on the potential of this source for muon colliders.
\bibliography{AMuonSourceBasedOnChannelingRadiation}

\begin{thebibliography}{11}%
\makeatletter
\providecommand \@ifxundefined [1]{%
 \@ifx{#1\undefined}
}%
\providecommand \@ifnum [1]{%
 \ifnum #1\expandafter \@firstoftwo
 \else \expandafter \@secondoftwo
 \fi
}%
\providecommand \@ifx [1]{%
 \ifx #1\expandafter \@firstoftwo
 \else \expandafter \@secondoftwo
 \fi
}%
\providecommand \natexlab [1]{#1}%
\providecommand \enquote  [1]{``#1''}%
\providecommand \bibnamefont  [1]{#1}%
\providecommand \bibfnamefont [1]{#1}%
\providecommand \citenamefont [1]{#1}%
\providecommand \href@noop [0]{\@secondoftwo}%
\providecommand \href [0]{\begingroup \@sanitize@url \@href}%
\providecommand \@href[1]{\@@startlink{#1}\@@href}%
\providecommand \@@href[1]{\endgroup#1\@@endlink}%
\providecommand \@sanitize@url [0]{\catcode `\\12\catcode `\$12\catcode
  `\&12\catcode `\#12\catcode `\^12\catcode `\_12\catcode `\%12\relax}%
\providecommand \@@startlink[1]{}%
\providecommand \@@endlink[0]{}%
\providecommand \url  [0]{\begingroup\@sanitize@url \@url }%
\providecommand \@url [1]{\endgroup\@href {#1}{\urlprefix }}%
\providecommand \urlprefix  [0]{URL }%
\providecommand \Eprint [0]{\href }%
\providecommand \doibase [0]{http://dx.doi.org/}%
\providecommand \selectlanguage [0]{\@gobble}%
\providecommand \bibinfo  [0]{\@secondoftwo}%
\providecommand \bibfield  [0]{\@secondoftwo}%
\providecommand \translation [1]{[#1]}%
\providecommand \BibitemOpen [0]{}%
\providecommand \bibitemStop [0]{}%
\providecommand \bibitemNoStop [0]{.\EOS\space}%
\providecommand \EOS [0]{\spacefactor3000\relax}%
\providecommand \BibitemShut  [1]{\csname bibitem#1\endcsname}%
\let\auto@bib@innerbib\@empty
\bibitem [{\citenamefont {Korol}\ \emph {et~al.}(2014)\citenamefont {Korol},
  \citenamefont {yov},\ and\ \citenamefont {Greiner}}]{channellingBook}%
  \BibitemOpen
  \bibfield  {author} {\bibinfo {author} {\bibfnamefont {A.}~\bibnamefont
  {Korol}}, \bibinfo {author} {\bibfnamefont {A.~S.}\ \bibnamefont {yov}}, \
  and\ \bibinfo {author} {\bibfnamefont {W.}~\bibnamefont {Greiner}},\
  }\href@noop {} {\emph {\bibinfo {title} {{Channeling and radiation in
  periodically bent crystals; 2nd ed.}}}},\ Springer Series on Atomic Optical
  and Plasma Physics\ (\bibinfo  {publisher} {Springer},\ \bibinfo {address}
  {Berlin},\ \bibinfo {year} {2014})\BibitemShut {NoStop}%
\bibitem [{\citenamefont {Schm{\"u}ser}\ \emph {et~al.}(2014)\citenamefont
  {Schm{\"u}ser}, \citenamefont {Dohlus}, \citenamefont {Rossbach},\ and\
  \citenamefont {Behrens}}]{undulatorTheory}%
  \BibitemOpen
  \bibfield  {author} {\bibinfo {author} {\bibfnamefont {P.}~\bibnamefont
  {Schm{\"u}ser}}, \bibinfo {author} {\bibfnamefont {M.}~\bibnamefont
  {Dohlus}}, \bibinfo {author} {\bibfnamefont {J.}~\bibnamefont {Rossbach}}, \
  and\ \bibinfo {author} {\bibfnamefont {C.}~\bibnamefont {Behrens}},\
  }\href@noop {} {\emph {\bibinfo {title} {{Free-electron lasers in the
  ultraviolet and X-ray regime: physical principles, experimental results,
  technical realization; 2nd ed.}}}},\ Springer Tracts in Modern Physics\
  (\bibinfo  {publisher} {Springer},\ \bibinfo {address} {Cham},\ \bibinfo
  {year} {2014})\BibitemShut {NoStop}%
\bibitem [{\citenamefont {Burkhardt}\ \emph {et~al.}(2002)\citenamefont
  {Burkhardt}, \citenamefont {Kelner},\ and\ \citenamefont
  {Kokoulin}}]{helmut}%
  \BibitemOpen
  \bibfield  {author} {\bibinfo {author} {\bibfnamefont {H.}~\bibnamefont
  {Burkhardt}}, \bibinfo {author} {\bibfnamefont {S.}~\bibnamefont {Kelner}}, \
  and\ \bibinfo {author} {\bibfnamefont {R.}~\bibnamefont {Kokoulin}},\
  }\href@noop {} {\emph {\bibinfo {title} {{Monte Carlo Generator for Muon Pair
  Production}}}},\ \bibinfo {type} {Tech. Rep.}\ \bibinfo {number}
  {CERN-SL-2002-016-AP}\ (\bibinfo  {institution} {CERN},\ \bibinfo {address}
  {Geneva},\ \bibinfo {year} {2002})\BibitemShut {NoStop}%
\bibitem [{\citenamefont {Clarke}(2004)}]{undulatorTheory2}%
  \BibitemOpen
  \bibfield  {author} {\bibinfo {author} {\bibfnamefont {J.}~\bibnamefont
  {Clarke}},\ }\href@noop {} {\emph {\bibinfo {title} {The science and
  technology of undulators and wigglers}}},\ Oxford series on synchrotron
  radiation\ (\bibinfo  {publisher} {Oxford University Press},\ \bibinfo
  {address} {Oxford},\ \bibinfo {year} {2004})\BibitemShut {NoStop}%
\bibitem [{\citenamefont {Litvinenko}\ \emph {et~al.}(2005)\citenamefont
  {Litvinenko}, \citenamefont {Ben-Zvi}, \citenamefont {Ahrens}, \citenamefont
  {Bai}, \citenamefont {Beebe-Wang}, \citenamefont {Blaskievicz}, \citenamefont
  {Brennan}, \citenamefont {Calaga}, \citenamefont {Chang}, \citenamefont
  {Fedotov}, \citenamefont {Fischer}, \citenamefont {Kayran}, \citenamefont
  {Kewisch}, \citenamefont {MacKay}, \citenamefont {Montag}, \citenamefont
  {Parker}, \citenamefont {Peggs}, \citenamefont {Ptitsyn}, \citenamefont
  {Roser}, \citenamefont {Ruggiero}, \citenamefont {Satogata}, \citenamefont
  {Surrow}, \citenamefont {Tepikian}, \citenamefont {Trbojevic}, \citenamefont
  {Yakimenko},\ and\ \citenamefont {Zhang}}]{eRHIC}%
  \BibitemOpen
  \bibfield  {author} {\bibinfo {author} {\bibfnamefont {V.}~\bibnamefont
  {Litvinenko}}, \bibinfo {author} {\bibfnamefont {I.}~\bibnamefont {Ben-Zvi}},
  \bibinfo {author} {\bibfnamefont {L.}~\bibnamefont {Ahrens}}, \bibinfo
  {author} {\bibfnamefont {M.}~\bibnamefont {Bai}}, \bibinfo {author}
  {\bibfnamefont {J.}~\bibnamefont {Beebe-Wang}}, \bibinfo {author}
  {\bibfnamefont {M.}~\bibnamefont {Blaskievicz}}, \bibinfo {author}
  {\bibfnamefont {J.}~\bibnamefont {Brennan}}, \bibinfo {author} {\bibfnamefont
  {R.}~\bibnamefont {Calaga}}, \bibinfo {author} {\bibfnamefont
  {X.}~\bibnamefont {Chang}}, \bibinfo {author} {\bibfnamefont
  {A.}~\bibnamefont {Fedotov}}, \bibinfo {author} {\bibfnamefont
  {W.}~\bibnamefont {Fischer}}, \bibinfo {author} {\bibfnamefont
  {D.}~\bibnamefont {Kayran}}, \bibinfo {author} {\bibfnamefont
  {J.}~\bibnamefont {Kewisch}}, \bibinfo {author} {\bibfnamefont
  {W.}~\bibnamefont {MacKay}}, \bibinfo {author} {\bibfnamefont
  {C.}~\bibnamefont {Montag}}, \bibinfo {author} {\bibfnamefont
  {B.}~\bibnamefont {Parker}}, \bibinfo {author} {\bibfnamefont
  {S.}~\bibnamefont {Peggs}}, \bibinfo {author} {\bibfnamefont
  {V.}~\bibnamefont {Ptitsyn}}, \bibinfo {author} {\bibfnamefont
  {T.}~\bibnamefont {Roser}}, \bibinfo {author} {\bibfnamefont
  {A.}~\bibnamefont {Ruggiero}}, \bibinfo {author} {\bibfnamefont
  {T.}~\bibnamefont {Satogata}}, \bibinfo {author} {\bibfnamefont
  {B.}~\bibnamefont {Surrow}}, \bibinfo {author} {\bibfnamefont
  {S.}~\bibnamefont {Tepikian}}, \bibinfo {author} {\bibfnamefont
  {D.}~\bibnamefont {Trbojevic}}, \bibinfo {author} {\bibfnamefont
  {V.}~\bibnamefont {Yakimenko}}, \ and\ \bibinfo {author} {\bibfnamefont
  {S.}~\bibnamefont {Zhang}},\ }in\ \href@noop {} {\emph {\bibinfo {booktitle}
  {Proceeding of 2005 Particle Accelerator Conference}}}\ (\bibinfo
  {publisher} {Piscatawaz, IEEE.},\ \bibinfo {year} {2005})\BibitemShut
  {NoStop}%
\bibitem [{\citenamefont {{LHeC Study Group}}(2012)}]{LHeC}%
  \BibitemOpen
  \bibfield  {author} {\bibinfo {author} {\bibnamefont {{LHeC Study Group}}},\
  }\href@noop {} {\bibfield  {journal} {\bibinfo  {journal} {J. Phys. G: Nucl.
  Part. Phys.}\ }\textbf {\bibinfo {volume} {39}},\ \bibinfo {pages} {075001}
  (\bibinfo {year} {2012})}\BibitemShut {NoStop}%
\bibitem [{\citenamefont {M{\"o}hl}(2006)}]{multipleScatteringMohl}%
  \BibitemOpen
  \bibfield  {author} {\bibinfo {author} {\bibfnamefont {D.}~\bibnamefont
  {M{\"o}hl}},\ }in\ \href@noop {} {\emph {\bibinfo {booktitle} {{CAS} - {CERN}
  Accelerator School: Intermediate Course on Accelerator Physics}}},\ \bibinfo
  {editor} {edited by\ \bibinfo {editor} {\bibfnamefont {D.}~\bibnamefont
  {Brandt}}}\ (\bibinfo  {publisher} {CERN},\ \bibinfo {address} {Geneva,
  Switzerland},\ \bibinfo {year} {2006})\ pp.\ \bibinfo {pages}
  {245--270}\BibitemShut {NoStop}%
\bibitem [{\citenamefont {Wolski}(2014)}]{wolski}%
  \BibitemOpen
  \bibfield  {author} {\bibinfo {author} {\bibfnamefont {A.}~\bibnamefont
  {Wolski}},\ }\href@noop {} {\emph {\bibinfo {title} {Beam dynamics in high
  energy particle accelerators}}}\ (\bibinfo  {publisher} {Imperial College
  Press},\ \bibinfo {address} {London},\ \bibinfo {year} {2014})\BibitemShut
  {NoStop}%
\bibitem [{ele()}]{electronicLossTable}%
  \BibitemOpen
  \href@noop {} {}\bibinfo {howpublished}
  {https://physics.nist.gov/PhysRefData/Star/Text/intro.html}\BibitemShut
  {NoStop}%
\bibitem [{\citenamefont {Barlow}(2011)}]{prism}%
  \BibitemOpen
  \bibfield  {author} {\bibinfo {author} {\bibfnamefont {R.}~\bibnamefont
  {Barlow}},\ }\href@noop {} {\bibfield  {journal} {\bibinfo  {journal}
  {Nuclear Physics B - Proceedings Supplements}\ }\textbf {\bibinfo {volume}
  {218}},\ \bibinfo {pages} {44 } (\bibinfo {year} {2011})},\ \bibinfo {note}
  {proceedings of the Eleventh International Workshop on Tau Lepton
  Physics}\BibitemShut {NoStop}%
\bibitem [{\citenamefont {Blackmore}(2015)}]{nustorm}%
  \BibitemOpen
  \bibfield  {author} {\bibinfo {author} {\bibfnamefont {V.}~\bibnamefont
  {Blackmore}},\ }\href@noop {} {\bibfield  {journal} {\bibinfo  {journal}
  {Nuclear and Particle Physics Proceedings}\ }\textbf {\bibinfo {volume}
  {265-266}},\ \bibinfo {pages} {205 } (\bibinfo {year} {2015})},\ \bibinfo
  {note} {proceedings of the Neutrino Oscillation Workshop}\BibitemShut
  {NoStop}%
\end{thebibliography}%
\end{document}